%% file: main.tex
\documentclass[acmsmall]{acmart}
\usepackage{etex}

\usepackage{subcaption} 

\acmConference[]{}

\usepackage{booktabs}   
\usepackage{subcaption} 
\usepackage[noend]{algpseudocode}

\author{Yuan Si}
\affiliation{%
  \institution{University of Waterloo}
  \city{Waterloo}
  \country{Canada}
  }
\email{yuan.si@uwaterloo.ca}

\author{Daming Li}
\affiliation{%
  \institution{Independent Researcher}
  \country{USA}
  }
\email{damingliyale22@gmail.com}

\author{Hanyuan Shi}
\affiliation{%
  \institution{Independent Researcher}
  \country{China}
  }
\email{shihanyuan1995@gmail.com}

\author{Jialu Zhang}
\authornote{Corresponding Author}
\affiliation{%
  \institution{University of Waterloo}
  \city{Waterloo}
  \country{Canada}
  }
\email{jialu.zhang@uwaterloo.ca}

\input{preamble}

\begin{document}

\title{ViScratch: Using Large Language Models and Gameplay Videos for Automated Feedback in Scratch}

\begin{abstract}

\input{secs/abs}
\end{abstract}

\maketitle

\input{secs/intro}
\input{secs/background}

\input{secs/merge}
\input{secs/systemdesign}

\input{secs/eval}
\input{secs/discussion}
\input{secs/related_work}

\input{secs/conclusion}

\newpage



\bibliographystyle{ACM-Reference-Format}
\bibliography{scratch}

\end{document}

%% file: preamble.tex
\usepackage{graphicx}
\usepackage{amsmath}

\let\emptyset\varnothing
\usepackage{xspace}
\usepackage{footnote}
\usepackage{amsfonts}
\usepackage{natbib}
\usepackage{hhline}
\usepackage{tabularx}
\usepackage{scratch3}
\usepackage{diagbox}
\usepackage{multirow}
\usepackage{setspace}
\usepackage{epsfig}
\usepackage{url}

\usepackage{breakurl}
\usepackage{booktabs}
\usepackage{tabularx}
\newcolumntype{Y}{>{\raggedright\arraybackslash}X}
\usepackage{tikz}
\usetikzlibrary{arrows.meta,positioning,shapes.geometric,shapes.misc,fit,backgrounds}

\usepackage{xcolor}
\usepackage{lipsum}
\usepackage{courier}
\usepackage{listings}
\usepackage{caption}
\usepackage{colortbl}
\usepackage{arydshln} 
\usepackage{makecell} 
\usepackage{tikz}
\usepackage{color}
\usepackage{rotating}

\usepackage{varwidth}
\usetikzlibrary{fit,arrows,calc,positioning,quotes}

\usepackage{pgfplots}

\usepackage{mathtools}
\usepackage{fancyvrb}
\usepackage{textcomp}

\usepackage{longtable}
\usepackage{pdflscape} 
\usepackage{threeparttable} 
\usepackage{array}     

\usepackage{stmaryrd}

\usepackage{algorithm}
\usepackage[noend]{algpseudocode}

\usepackage{alltt}

\tikzset{
  -|-/.style={
    to path={
      (\tikztostart) -| ($(\tikztostart)!#1!(\tikztotarget)$) |- (\tikztotarget)
      \tikztonodes
    }
  },
  -|-/.default=0.5,
  |-|/.style={
    to path={
      (\tikztostart) |- ($(\tikztostart)!#1!(\tikztotarget)$) -| (\tikztotarget)
      \tikztonodes
    }
  },
  |-|/.default=0.5
}

\usetikzlibrary{shapes.geometric}

\long\def\com#1{}

\newcommand{\app}{\texttt{ViScratch}\xspace}

\newcommand{\para}[1]{\smallskip\noindent {\bf #1}}

\newcommand{\squishlist}{
   \begin{list}{$\bullet$}
    { \setlength{\itemsep}{0pt}      \setlength{\parsep}{3pt}
      \setlength{\topsep}{3pt}       \setlength{\partopsep}{0pt}
      \setlength{\leftmargin}{3.5mm} \setlength{\labelwidth}{1em}
      \setlength{\labelsep}{0.5em} }
}

\newcommand{\squishend}{
    \end{list}  }

\def\BibTeX{{\rm B\kern-.05em{\sc i\kern-.025em b}\kern-.08em
    T\kern-.1667em\lower.7ex\hbox{E}\kern-.125emX}}

%% file: secs/abs.tex
Block-based programming environments such as Scratch are increasingly popular in programming education, in particular for young learners. While the use of blocks helps prevent syntax errors, semantic bugs remain common and difficult to debug.  Existing tools for Scratch debugging rely heavily on predefined rules or user manual inputs, and crucially, they ignore the platform’s inherently visual nature.

We introduce \app, the first multimodal feedback generation system for Scratch that leverages both the project’s block code and its generated gameplay video to diagnose and repair bugs. \app uses a two-stage pipeline: a vision-language model first aligns visual symptoms with code structure to identify a single critical issue, then proposes minimal, abstract syntax tree level repairs that are verified via execution in the Scratch virtual machine.

We evaluate \app on a set of real-world Scratch projects against state-of-the-art LLM-based tools and human testers. Results show that gameplay video is a crucial debugging signal: \app substantially outperforms prior tools in both bug identification and repair quality, 
even without access to project descriptions or goals. 
This work demonstrates that video can serve as a first-class specification in visual programming environments, opening new directions for LLM-based debugging beyond symbolic code alone.

%% file: secs/intro.tex
\section{Introduction}
\label{sec:intro}

Block-based programming environments like Scratch~\cite{resnick2009scratch, maloney2010scratch} have become foundational in introductory computing education, particularly for young learners. Scratch alone has over 140 million registered users worldwide to date\footnote{\url{https://annualreport.scratchfoundation.org}}, enabling users to create games, animations, and interactive stories by stacking together visual blocks rather than writing textual code. This design eliminates syntax errors by construction, but semantic bugs still abound. Learners often encounter unexpected behaviors, but without explicit error messages, many remain unaware that a bug exists~\cite{fradrich2020commonbugs}, let alone how to fix it. Online forums are flooded with such questions\footnote{\url{https://scratch.mit.edu/discuss/}}, revealing a critical need: scalable, automated feedback that can help students understand and resolve bugs in their  projects.

A growing body of research~\cite{deiner2024nuzzlebug, STRIJBOL2024101617, schweikl2025repurr} has proposed tools to support Scratch debugging, including interactive debuggers with stepping and breakpoints, and search-based program repair engines. These tools, however, heavily rely on user-driven interpretations or hand-crafted rules. Similarly, automated feedback systems~\cite{price2017isnap, fein2022catnip} for block-based programming often depend on instructor-authored test suites or predefined patterns, limiting their applicability to open-ended learner projects.

We argue for a fundamentally different perspective: Scratch’s visual output should be treated as a first-class debugging signal rather than a secondary artifact. Scratch programs are inherently visual, and correctness is typically judged by what learners observe rather than by textual output or internal traces. Gameplay videos, generated during execution, capture learner intent and expose perceptual bugs such as flickering sprites, missed collisions, and animation glitches that cannot be diagnosed from code alone. Video playback is therefore not a weaker proxy, but the de facto oracle used in classrooms.

At the same time, LLMs have shown strong potential in traditional software engineering tasks such as bug detection and program repair~\cite{chen2021codex}. Yet their application to block-based programming is still nascent~\cite{DrugaOtero2023,GrieblFOFJ2023}, and none, to our knowledge, incorporate visual runtime signals.
Recent AI copilots for Scratch~\cite{chen2024chatscratch, chen2025mindscratch, druga2025scratch} primarily focus on enhancing creativity and planning, not debugging. Furthermore, LLMs struggle with editing raw JSON representations of Scratch projects, often producing invalid or unexecutable outputs. We hypothesize that using these gameplay videos as input to an LLM can unlock a new frontier in automated debugging for visual programming. 

Given these motivations, we present \app, the first LLM-powered debugging system for Scratch that uses both code and gameplay video as multimodal input. \app features a two-stage architecture: a diagnosis stage uses a vision-language model to align runtime video behavior with the project’s abstract syntax tree (AST) and identify the root cause of failure; a repair stage then applies minimal AST edits and verifies them through execution in the Scratch virtual machine. To support reliable repair, we normalize projects, enforce minimal-edit policies, and use an iterative update loop that integrates learner feedback and retry history.

In summary, this paper makes the following contributions:
\begin{itemize}
    \item We design and implement \app, a multimodal feedback generation system that diagnoses and repairs Scratch programs using both block code and gameplay video as input.
    \item We ensure the program fix quality and reliability of the system by combining  abstract syntax tree operations, minimal edit policies, and virtual machine validation.
    \item We conduct an empirical study of \app's performance in bug identification and fixing on real-world Scratch projects, benchmarking it against a state-of-the-art LLM baseline (without video) and human learners.
\end{itemize}

Overall, our results show that gameplay video is an indispensable debugging signal in Scratch. \app not only outperforms baselines on bug identification and repair success, but also provides actionable, verifiable fixes, without requiring access to project descriptions or reference solutions. This work positions video as a new signal surface for program reasoning, opening up promising directions in multimodal software engineering.

%% file: secs/background.tex
\section{Background}
\label{sec:background} 

\para{The Scratch Interface and Semantics.}
Scratch is a block-based language where learners compose scripts by stacking visual blocks.
Its programming interface is divided into four main areas: the block palette (left) with categorized building blocks (e.g., motion, control), the scripting workspace (center) for composing behaviors, the stage (upper right) for real-time visual feedback, and a panel (lower right) for managing sprites and backdrops. This layout enables learners to experiment interactively and connect logic with behavior through immediate feedback. Behind the scenes, these programs are serialized into a structured intermediate representation (JSON) that encodes other necessary auxiliary parts to help run the program such as media assets. This intermediate representation feeds into the Scratch Virtual Machine (VM), which executes each script as a thread in an event-driven model. This design emphasizes robustness: runtime exceptions are silently ignored, allowing creative freedom but leaving learners unaware of many bugs.

\begin{figure}
    \centering
    \includegraphics[width=\linewidth]{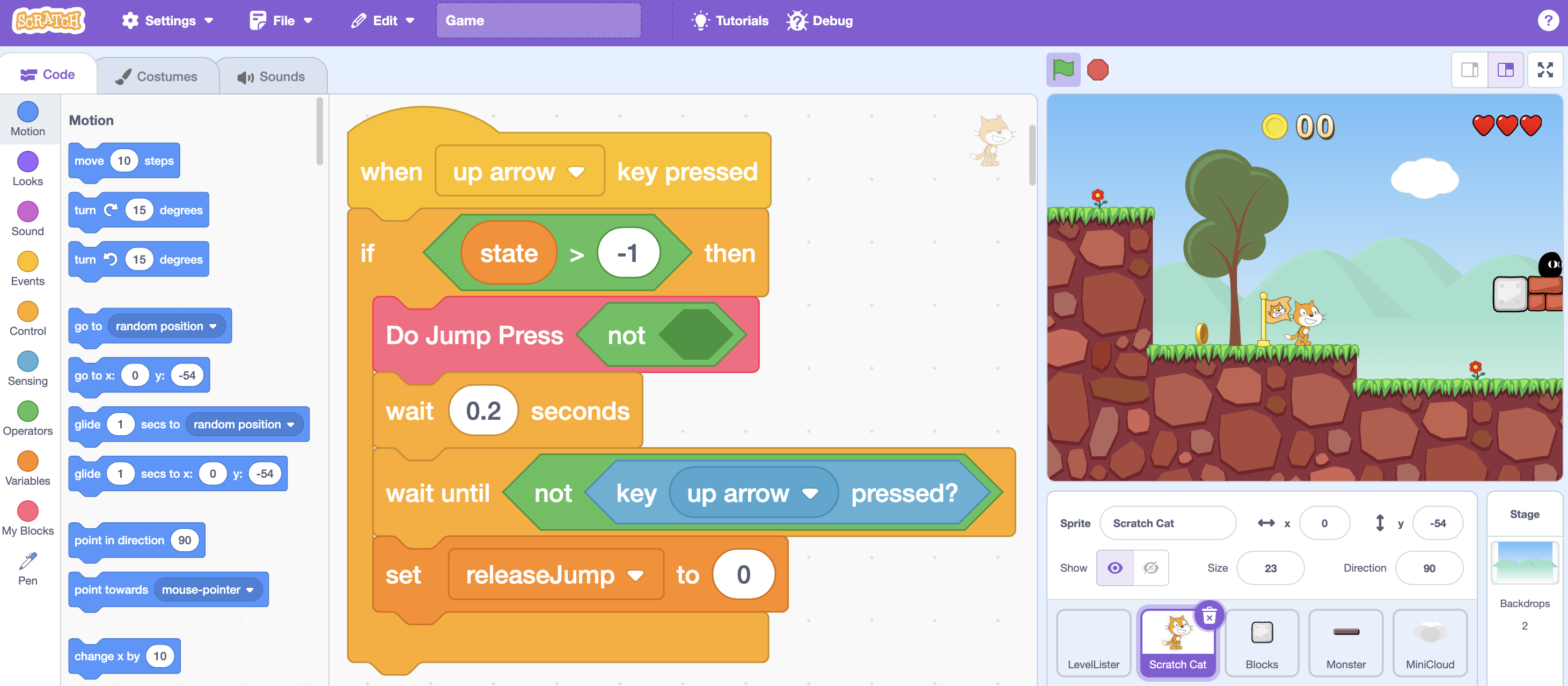}
    \caption{An example Scratch project simulating a Mario game, source: \url{https://scratch.mit.edu/projects/10118230/}}
    \label{fig:GUI}
\end{figure}

\para{Why the Video Matters.}
What sets Scratch apart from other imperative languages is that its correctness is often visual. Figure~\ref{fig:GUI} shows a Scratch project simulating a Mario game: users define logic using blocks (left), and observe behavior through the stage (right). Bugs are perceptual, a jump failing to trigger, a coin not disappearing, a sprite flickering, and these issues rarely manifest in left-hand blocks. Instead, they are only visible in the final rendered gameplay video.

\para{\app: Feedback Without Disruption.}
\app integrates seamlessly into the native Scratch workflow without altering how learners build or execute programs. It passively observes two artifacts already produced during normal use: the visual code blocks and the rendered gameplay video. When a learner encounters unexpected behavior, such as a sprite failing to react or an animation glitch, \app can be invoked to diagnose the issue and suggest a fix. Until then, it remains silent and external, preserving the learner’s experience.
This passive design is pedagogically critical. Novice programmers thrive on unstructured exploration and visual feedback. Tools that inject chat-based suggestions or require instrumentation risk disrupting this flow and overwhelming the user. By treating the gameplay video as the primary debugging surface and activating only when explicitly invoked, \app provides assistance while fully preserving learner agency and creative momentum.

%% file: secs/merge.tex
\section{Bug Patterns in Scratch: The Case for Video as a Debugging Signal}
\label{sec:merge}

Scratch programs unfold through visual dynamics: sprite movement, layering, timing, and visual effects. Unlike traditional programming environments where correctness is defined by symbolic state or textual output, Scratch correctness is inherently perceptual. Learners judge a program by what they \emph{see}. This raises a fundamental question: \emph{what if the primary debugging surface is not a trace, but a video?}

Our key insight is that in Scratch, \emph{gameplay video is not auxiliary, it is necessary}. Many impactful bugs manifest only at runtime through visual symptoms, such as flickering, desynchronization, layering glitches, which are difficult to discover in source code or event traces. To ground this claim, we present a taxonomy of bugs derived from 50 real-world Scratch projects\footnote{\url{https://scratch.mit.edu/discuss}} and show that perceptual cues are central to debugging in visual programming environments.

\subsection{Motivating Examples: When Analyzing Code Isn’t Enough}

We begin with three real projects that illustrate why traditional text-based debugging approaches may fail in Scratch.

\para{Example 1: Strobe Flicker.}

\begin{figure}[H]
\begin{center}
\begin{tikzpicture}[baseline=(current bounding box.south)]
  \node[scale=0.7] (left) at (0.8,0) {
    \begin{scratch}
      \blockinit{when receiving \selectmenu{show\_force}}
      \blockrepeat{forever}
      {
        \blockcontrol{wait \ovalnum{0.01} seconds}
        \blocklook{show}
      }
    \end{scratch}
  };
  \node[scale=0.7] (right) at (7.6,0) {
    \begin{scratch}
      \blockinit{when receiving \selectmenu{hide\_force}}
      \blockrepeat{forever}
      {
        \blockcontrol{wait \ovalnum{0.01} seconds}
        \blocklook{hide}
      }
    \end{scratch}
  };
  \draw[<->,red,ultra thick] (1,0.5) -- (5.9,0.5)
    node[midway,below,red,font=\bfseries] {Concurrent};
\end{tikzpicture}
\end{center}
\caption{Strobe flicker caused by two concurrent broadcast handlers that show and hide the sprite every 0.01s in an alternating manner, such that frame switching at 100 fps produces rapid visual flickering.}
\label{fig:strobe-flicker}
\end{figure}
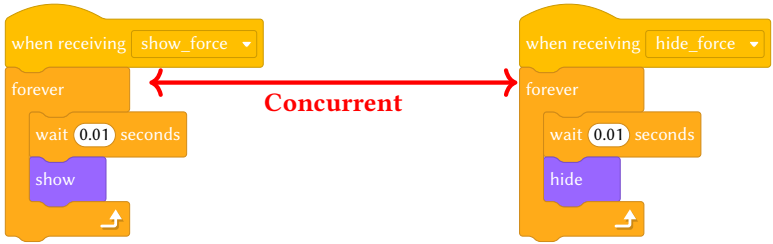

Project description: Figure~\ref{fig:strobe-flicker} shows the block code underlying a sprite which is expected to appear when the user presses the \texttt{s} key (triggering \texttt{show\_force}) and disappear when the \texttt{h} key is pressed (triggering \texttt{hide\_force}).

In this example, the two scripts run at the same time. One listens for a \texttt{show\_force} signal and displays the sprite at extremely fast frequency, while the other listens for a \texttt{hide\_force} signal and hides the sprite at the same frequency. Both scripts are correct when inspected separately. However, their concurrent execution causes a rapid flickering effect due to frame switching that is hard to be identified in code alone, but is rendered clearly in video. Traditional debugging tools may struggle to detect it due to lack of visual signal input. 
This case highlights why debugging support in Scratch should incorporate a richer set of signals due to its inherently visual characteristics.

\para{Example 2: Race Condition.}

\begin{figure}[H]
\begin{center}
\begin{tikzpicture}[baseline=(current bounding box.south)]
  \node[scale=0.7] (left) at (0.8,0) {
    \begin{scratch}
      \blockinit{when receiving \selectmenu{start}}
      \blockrepeat{forever}
      {
      \blockrepeat{if \ovalnum{cat} touches \ovalnum{bat}}
      {
        \blocklook{change position-X by \ovalnum{-10}}
      }}
    \end{scratch}
  };
  \node[scale=0.7] (right) at (7.6,0) {
    \begin{scratch}
      \blockinit{when receiving \selectmenu{start}}
      \blockrepeat{forever}
      {
      \blockrepeat{if \ovalnum{cat} touches \ovalnum{bat}}
      {
        \blocklook{change score by \ovalnum{-1}}
      }
      }
    \end{scratch}
  };
  \draw[<->,red,ultra thick] (2.6,-0.33) -- (6,0.1)
    node[midway,below,red,font=\bfseries] {Racing};
\end{tikzpicture}
\end{center}
\caption{Race condition caused by splitting sprite position update and score change into two concurrent forever loops in separate code blocks. With the \texttt{start} signal triggered, if \texttt{cat} touches \texttt{bat}, \texttt{cat}'s position is shifted by -10 (executed by the left block) and the score is decreased by 1 (executed in the right block). Here the score fails to update if the position update code is executed first.}
\label{fig:cat-bat-touch}
\end{figure}
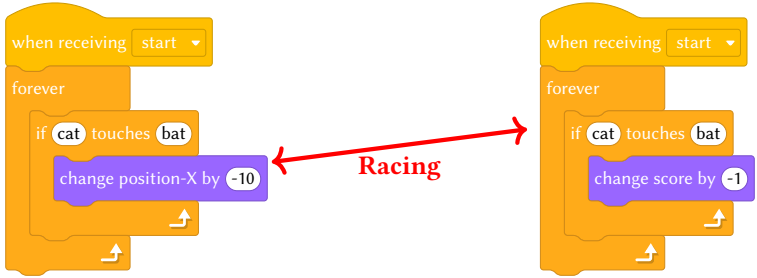

Project description: Figure~\ref{fig:cat-bat-touch} shows the block code of a game where the player controls a \texttt{cat} sprite. If the \texttt{cat} touches a \texttt{bat} sprite, \texttt{cat} is expected to bounce back and decrease the score by 1.

Because Scratch runs the two position and score updating scripts as independent threads, splitting the logic over two \texttt{forever} loops introduces a race condition. If the position update script is executed first, \texttt{cat} will no longer be able to touch \texttt{bat} when the score update script runs, causing the score change statement to be skipped, even if the logic appears correct in both scripts.
This bug proves to be very difficult to identified from static code inspection alone. In contrast, the gameplay video reveals the problem easily: \texttt{cat} is bounced back upon touching \texttt{bat} as expected, but the score remains unchanged. 

\para{Example 3: Duplicate Count.}

Project description: Figure~\ref{fig:duplicate-count} shows the block code of a game where the player clicks cat sprites to catch them. Once a cat is caught, count value should be increased by 1. There are $N$ cats in total, all sharing the same script as shown below:

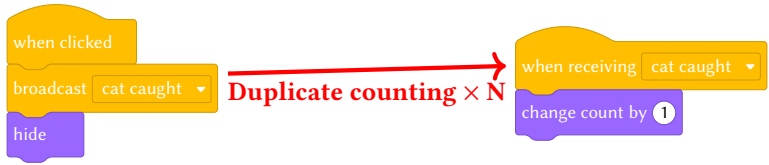
\begin{figure}[H]
\begin{center}
\begin{tikzpicture}[baseline=(current bounding box.south)]
    \node[scale=0.7] (left) at (0.76,0) {
    \begin{scratch}
      \blockinit{when clicked}
      \blockevent{broadcast \selectmenu{cat caught}}
      \blocklook{hide}
    \end{scratch}
  };
  \node[scale=0.7] (left) at (7.8,0) {
    \begin{scratch}
      \blockinit{when receiving \selectmenu{cat caught}}
      \blocklook{change count by \ovalnum{1}}
    \end{scratch}
  };
  \draw[->,red,ultra thick] (2.3,0.1) -- (6,0.2)
    node[midway,below,red,font=\bfseries] {Duplicate counting $\times$ N};
\end{tikzpicture}
\end{center}
\caption{Duplicate counting caused by global broadcast in a Cat Catcher game. Clicking a sprite broadcasts the "cat caught" message (left block). Every sprite that listens to this message executes changing count value by 1 (right block). With $N$ listeners, a single click erroneously yields an increase of $N$ in count.}
\label{fig:duplicate-count}
\end{figure}

Because Scratch uses a global scope for broadcasts, placing the count updating script under the ``cat caught'' message in listener's code causes every listening sprite to increase the count value. Thus one click fans out to $N$ increments. Each script looks correct in isolation, which makes the bug easy to be overlooked during code inspection.
The gameplay video reveals the error immediately: the count value increases by $N$ instead of 1 after a single click. One can identify this bug from the gameplay video by checking whether the increments triggered by the broadcast matches the expected count update from a single click. Again, this case highlights why debugging support in Scratch needs to go beyond code inspection. Watching the gameplay video makes detecting and resolving such issues much easier.

\subsection{Understanding Scratch Bug Patterns}

\para{Methodology.}
We present our study on 50 real-world bugs on the official Scratch forum \footnote{\url{https://scratch.mit.edu/discuss}}. We describe the common patterns of Scratch bugs and give some concrete
examples. We further discuss the implications that drive the design of \app.

We collected 50 user-reported Scratch issues from official Scratch forum, the largest Q\&A forum where Scratch-related issues are commonly reported. We identified Scratch bugs by looking for issues in
which users reported bugs in the code. For example,
in one post \footnote{\url{https://scratch.mit.edu/discuss/topic/414517}}, the question was ``cloned sprite is responding to original sprite's event which is not what I want to see'' and the correct answer was ``use clone-index variable to identify.'' Specifically, we selected posts that contain the
keywords such as ``I don't know'', ``confused'',  etc. in either the question or
any of the answers. 
We excluded unanswered posts to ensure that each bug’s ground truth could be validated and classified.
We manually analyzed each post to understand the Scratch bug,
including the bug pattern and the source-code root cause.

\begin{table}[t]
\centering
\caption{Bug patterns and whether video is essential for diagnosis. Code-visible: bugs that can be found by inspecting blocks or event handlers. Video-required: symptoms need to be manifested in gameplay video. }
\begin{tabular}{lcc}
\toprule
\textbf{Bug Pattern} & \textbf{Video-required} & \textbf{Code-visible} \\
\midrule
Missing clone operation        & 10 & 0 \\
Recursive cloning             & 8 & 0 \\
Missing termination condition & 4 & 1 \\
Wrong parameter values        & 3 & 1 \\
Wrong logic inside condition  & 1 & 4 \\
Missing loop sensing          & 1 & 3 \\
Message never received        & 1 & 2 \\
Forever inside loop           & 1 & 1 \\
Custom block with termination & 1 & 0 \\
Stuttering movement           & 1 & 0 \\
Wrong comparisons/thresholds  & 1 & 0 \\
Missing x interaction script          & 1 & 0 \\
Missing backdrop switch       & 1 & 0 \\
No working scripts            & 1 & 0 \\
Expression as touch/colour    & 0 & 1 \\
Custom block with forever     & 0 & 1 \\
Comparing literals            & 0 & 1 \\
\midrule
\textbf{Total}                & 35 & 15 \\
\bottomrule
\end{tabular}
\label{tab:video-bugs}
\end{table}

\subsection{Findings}

\para{Finding 1: Clone-related bugs are inherently perceptual.}

Across 17 cases involving clone behaviors (initialization, deletion, recursion), \textbf{100\%} required video. These account for \textbf{34\%} of the dataset, making clones the most video-dependent class. The decisive cues, exponential sprite growth, persistent overlaps, or vanishing menus, were visible only in gameplay video, never in block inspection. Video is therefore indispensable for diagnosing clone lifecycle failures.

\para{Finding 2: Code-visible bugs still gain from video.}

For control-flow and logic bugs, \textbf{67\% (10/15)} could be diagnosed from static inspection alone. Yet video consistently revealed the \emph{runtime manifestations}, variables jumping unpredictably, characters sinking instantly, missed collisions, that made the bug’s impact clear. Code showed \emph{what} was wrong; video showed \emph{how disruptive it felt}. Even in ``code-visible'' categories, video enriched debugging by exposing perceptual consequences.

\para{Finding 3: Video clarifies severity at scale.}

Overall, \textbf{70\% (35/50)} of all bugs required video as the decisive signal. In the remaining \textbf{30\%}, where code sufficed, gameplay video enhanced triage by illustrating how seemingly minor mistakes (e.g., thresholds off by one) escalated into disruptive gameplay. Thus, video’s role is not binary but gradient, essential for two-thirds, clarifying for the rest, making it a universal debugging aid.

\para{Insight.} Debugging in visual programming environments cannot rely on block inspection alone: video is a first-class signal, indispensable in most cases and impactful in the rest. Our survey shows that video is not an auxiliary artifact but showing the ground truth of correctness. Many impactful bugs are perceptual at their core, even when the underlying fix is within a single block. These findings not only characterize recurring failure modes in Scratch but also provide direct guidance for tool builders: they identify which bugs require perceptual evidence, where automation is feasible, and how video must be synchronized with program structure. Building on these insights, we designed \app, which combines code and gameplay video to diagnose and repair Scratch programs.

%% file: secs/systemdesign.tex
\section{System Design}
\label{sec:design}
\begin{figure}[t]
    \centering
    \includegraphics[width=\textwidth]{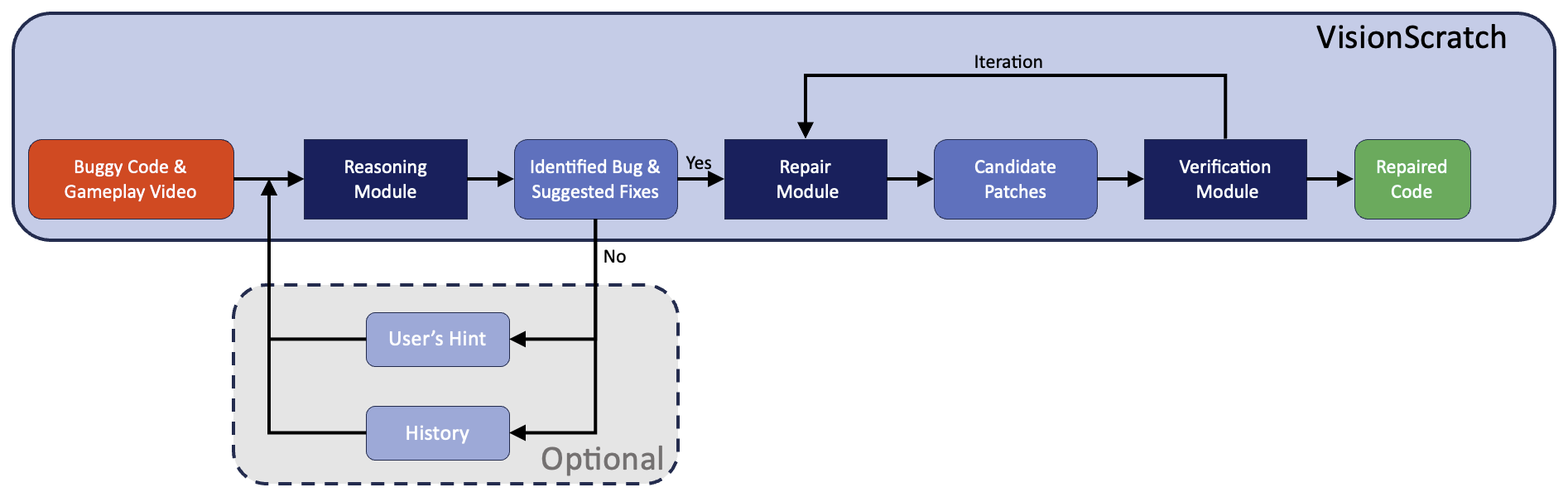}
    \caption{\app system pipeline. Given a learner’s Scratch buggy code and its corresponding gameplay video, the reasoning module jointly analyzes code and visual behavior to identify the root cause of failure and propose candidate fixes. When applicable, optional user hints and retry history are consulted to refine the diagnosis. The selected fix is passed to the repair module, which applies edits and produces candidate patches. These patches are executed and verified within the verification module. If the verification step fails, the system keeps iterating by updating history and revisiting the repair process. Once verification succeeds, the system outputs the repaired code.}
    \label{fig:system-flow}
\end{figure}

This section introduces the overall architecture and key components of the \app system, as shown in Figure~\ref{fig:system-flow}. 
The system inputs include the learner’s Scratch buggy code (visual code blocks, intermediate representation (JSON) and media assets) and the produced gameplay video. With these four types of inputs, the reasoning module identifies the bug in the project and proposes several candidate fixes. The repair module then automatically implements the selected fix, followed by verification module checking its effectiveness.
This end-to-end pipeline synthesizes analysis of Scratch source code, gameplay video interpretation, and LLM-based reasoning to automate debugging and repair. Below, we define the key terminology and outline the system flow and algorithm.

\begin{algorithm}[t]
\caption{\app: Diagnose-Repair-Verify}
\label{alg:visionscratch}
\begin{algorithmic}[1]
\Require buggy project $B$ (blocks, JSON, media assets), gameplay video $V$, retry history $H$(optional), user hint $UH$(optional), iteration cap $K$
\State $B' \gets \textsc{Normalize}(B)$; \quad $V' \gets \textsc{Normalize}(V)$
\Repeat
  \State $(\textit{bug}, \textit{fixes}) \gets \textsc{ReasoningModule}(V', B', H, UH)$ 
  \State \textsc{ShowToUser}$(\textit{bug}, \textit{fixes})$
  \If{\textsc{UserSatisfied}() = \textsc{False}}
     \State $UH \gets \textsc{UserUpdateHint}(UH)$
     \State $H \gets \textsc{Update}(H,\textit{bug},\textit{fixes},UH)$
  \EndIf
\Until{\textsc{UserSatisfied()} = \textsc{True}}
\State $\textit{fix} \gets \textsc{UserSelectFix}(\textit{fixes})$ 
\State $log \gets \emptyset$
\For{$k = 1 \dots K$} 
  \State $\textit{patch} \gets \textsc{RepairModule}(\textit{fix}, B', log)$
  \State $Repaired\_B \gets \textsc{AssembleProject}(B,\textit{patch})$
  \State $(\textit{verdict},log) \gets \textsc{VerificationModule}(Repaired\_B)$
  \If{$\textit{verdict} = \textsc{Pass}$} \Return $Repaired\_B$
  \Else \text{ }$H \gets \textsc{Update}(H,\textit{fix},\textit{patch},\textit{log})$
  \EndIf
\EndFor
\State \Return $Repaired\_B$
\end{algorithmic}

\end{algorithm}

\subsection{Architecture Overview}
The architecture of \app is shown in Figure~\ref{fig:system-flow}. The system operates in a three-stage loop of reasoning, repair, and verification over the learner’s buggy Scratch project (blocks, JSON, and media assets) together with its gameplay video. The reasoning module jointly interprets static structure and dynamic behavior to identify likely root causes and propose candidate fixes, optionally refined by user hints and retry history. The repair module applies selected fixes through AST-level edits, while the verification module executes the patched project in the Scratch virtual machine to check behavior. Failed patches trigger further iterations that integrate prior outcomes to improve results. Both reasoning and repair rely on Google’s Gemini 2.5 Pro model~\footnote{\url{https://ai.google.dev/gemini-api/docs/models##gemini-2.5-pro}}~\cite{Gemini25_techreport_2025}, enabling \app to unify gameplay video, program structure, and LLM-based reasoning in a tightly coupled pipeline. 
Algorithm~\ref{alg:visionscratch} illustrates \app’s core loop: it first interprets the buggy Scratch project, then suggests likely root cause and possible fixes to the user. If the user is not satisfied, the system refines its understanding using updated hints and history. Once a fix is selected, the system enters a bounded repair-and-verify loop. It iteratively synthesizes a patch, runs the program, and checks correctness. \app terminates when a passing fix is found or attempts are exhausted.

\subsection{Input and Representation}

\para{Scratch Project.} 
Scratch encodes projects using the \texttt{.sb3} format, a ZIP archive containing visual code blocks, intermediate representation (JSON) and media assets. \app parses JSON into an abstract syntax tree (AST), where nodes represent sprites and stages, edges capture block sequences, and symbol tables track variables and broadcasts. This structured representation supports targeted edits such as block insertion, replacement, and deletion during repair.

\para{Gameplay Video.} 
The video captures the project’s runtime behavior, exposing visual symptoms such as flickering, occlusion, or desynchronization. \app normalizes the video (FPS, resolution, color) and relies on a multimodal LLM to jointly interpret it with the project code, without requiring specialized vision models.

\para{User Hint and History.} 
Learners may optionally provide behavioral hints in natural language or reject prior fixes. \app additionally records past failed attempts. These inputs help the reasoning module refine diagnosis, eliminate redundancy, and guide future iterations.

\subsection{Reasoning Module}

The reasoning module in \app identifies major program flaws by jointly reasoning over code and video with a multimodal LLM. It begins with behavior mapping, where the model interprets sprite behaviors such as appearance, disappearance, motion, and interactions from the gameplay video and aligns them with abstract syntax tree (AST) blocks to assess whether the observed dynamics are explained by the code. It then performs defect identification by isolating the most critical flaw, such as a missing broadcast, a stale variable, or a nonresponsive collision condition, based on misalignments between visual outcomes and code logic. Finally, the system proposes two to three distinct repair strategies, each articulating a feasible corrective action (for example, "add a broadcast on collision" or "update score directly"), offering learners meaningful alternatives.
If a previous fix has been rejected, \app consults its interaction history to avoid making the same mistake and propose new directions. To guide diagnosis, the LLM is primed with a curated taxonomy of common Scratch bugs (such as block misuse, missing messages, or faulty loops), helping it map runtime symptoms to canonical error patterns and improve the reliability of reasoning.


\para{Prompt Engineering.}
The goal of prompt engineering in the reasoning module is to design a tightly scoped prompt that enforces conciseness, precision, and edit safety. As shown in Figure~\ref{fig:prompt_reasoning}, the prompt directs the LLM to align video playback with project buggy code, identify exactly one primary bug, and propose fix suggestions. A hint-based prompting strategy further strengthens the minimal edit principle by predefining common Scratch mistakes and canonical repair templates, thereby improving the quality, reliability, and stability of outputs. Shown in later evaluation, this design yields verifiable edits directly applicable to learners’ projects.

\begin{figure}[h!]
\centering
\begin{minipage}[t]{\linewidth}
\ttfamily\footnotesize
\textbf{Step 1 — Diagnosis (Buggy Code + Video)}\\
\textit{Inputs/Goal:} Blocks, JSON, media assets, gameplay video, latest feedback/history. 
Silent reasoning to locate the single most critical flaw.\\
\textit{Tasks:} (1) Enumerate observed sprite behaviors and cross-check with Blocks, JSON and media assets; 
(2) identify the primary bug; (3) propose 2–3 concise fixes.\\
\textit{Strict Output:} exactly 1–2 lines\\
\quad 1) \texttt{Bug description: <brief>}\\
\quad 2) \texttt{Fix options: A-<fix>, B-<fix>, C-<fix or omit>}\\
\textit{Guidelines:} Align video with code; detect missing/extra logic; adapt to feedback/history.\\
\textit{Bug Taxonomy:} \{block misuse/order/value; orphaned code; event conflicts; 
loops/branches/init; Scratch-specific: broadcast mismatch, clone lifecycle, layering/visibility, 
collision/bounds, scene transitions, audio–visual timing\}.\\
\end{minipage}
\caption{Prompt for reasoning module: aligning video and code to identify the bug and propose fix options in \app's diagnosis stage.}
\label{fig:prompt_reasoning}
\end{figure}

\subsection{Repair Module}

The repair module in \app takes the diagnosis as input and prompts the LLM to generate a precise patch to the buggy code. Two constraints guide this process. First, \app only modifies the script(s) directly tied to the diagnosed issue. Multiple edits are permitted only if absolutely necessary.
Second, each instruction must specify the target sprite, script location, and block-level edit (insert, replace, or delete).
The instructions are parsed into structured AST edits. The code modification component applies these edits by locating nodes, performing changes, and preserving link integrity. The result is a repaired AST, which is then repackaged into a new \texttt{.sb3} file. 

\para{Prompt Engineering.}
The goal of prompt engineering in the repair module is to translate the diagnosis into minimal, auditable JSON edits. In particular, we restrict the model to produce atomic block-level modifications rather than a large portion of code rewrite. By constraining both reasoning and output to strict formats (e.g., ``exactly 1–2 lines''). Figure~\ref{fig:prompt_repair} illustrates the prompt.

\begin{figure}[H]
\centering
\begin{minipage}[t]{\linewidth}
\ttfamily\footnotesize
\textbf{Step 2 — Patch (Atomic JSON Edits)}\\
\textit{Inputs/Goal:} Diagnosis (Step~1) + chosen fix; normalized buggy code, log. 
Apply the smallest safe JSON edits; modify only required sprites/scripts.\\
\textit{Allowed (priority):} (1) tweak literal/operator; (2) insert minimal block; 
(3) replace block with peer; (4) add missing hats; (5) delete harmful block.\\
\textit{Disallowed:} new features, new assets, wholesale rewrites, unrelated reorderings.\\
\textit{Anchoring:} Cite opcodes/anchors (\texttt{after}, \texttt{before}, \texttt{inside}, \texttt{wrap}); cross-sprite edits only if necessary.\\
\textit{Strict Output Format:} 1–3 lines; each line: \texttt{Step k: <sprite> — <edit>} 
($\leq$20 words).\\
\end{minipage}
\caption{Prompt for repair module: enforcing JSON edits in \app's repair stage.}
\label{fig:prompt_repair}
\end{figure}

\subsection{Verification Module}

The verification module loads the repaired project  and checks whether the defect is resolved. For example, if ``clicking a sprite should increase the score,'' the verifier checks if the score indeed changes after the clicking. If verification fails, the error is recorded in history, and the pipeline iteratively re-enters the repair stage to generate a new patched program. Empirically, we find that in our evaluation three or fewer iterations often suffice to repair common Scratch bugs.

\para{Summary.} By combining video evidence, static code analysis, and LLM reasoning, \app closes the loop from defect detection to repair. It generates targeted fixes, applies minimal edits, and verifies repairs in the Scratch VM, providing strong support for automated tutoring in programming education.

%% file: secs/eval.tex
\section{Evaluation}
\label{sec:eval}

To understand how \app identifies and fixes bugs in Scratch programs using the generated videos, we answer the following research questions:

\begin{itemize}
    \item RQ1: Are the program-generated videos on the platform useful signals for debugging?
   \item RQ2: How well can \app identify and fix program bugs?
\end{itemize}

\begin{table}[h]
  \centering
  \begin{tabularx}{\linewidth}{@{} l X l @{}}
    \toprule
    \textbf{Project Name} & \textbf{Bug Description} & \textbf{Difficulty Level} \\
    \midrule
    1. Cat Movement & Cat's movement is jerky - each step is too large & easy \\
    2. Cat Hide/Show Toggle & Cat keeps flashing - Hide/Show scripts' racing & easy \\
    3. Cat Catcher I & Count value never updates - variable name mismatch & easy \\
    4. Cat Catcher II & Count value increases N per collection - global broadcast triggers all score-update scripts & easy \\
    5. Cat Catcher III & Count value carries over - no reset script & easy \\
    6. Cat/Bat Collision I& Score goes below 0 - no terminating condition & easy \\
    7. Cat/Bat Collision II & Score never updates - bounce back/touch check scripts racing & medium\\
    8. Cat Catcher IV & Count value never updates - broadcast name mismatch & medium \\
    9. Apple Collector & Score accrues continuously under stuck contact - no apple collect script & hard \\
    10. Bugs Eater & Bat falls - wrong edge color picked & hard \\

    \bottomrule
  \end{tabularx}
  \caption{Overview of the experiment's dataset, with each row corresponding to one Scratch project with their bug description, and difficulty level (easy/medium/hard).}
  \label{fig:Dataset Descr.}
\end{table}

\subsection{Experimental Setup}
\label{sec:exp-setup}

\begin{figure}[t!]
    \centering
    \includegraphics[width=\linewidth]{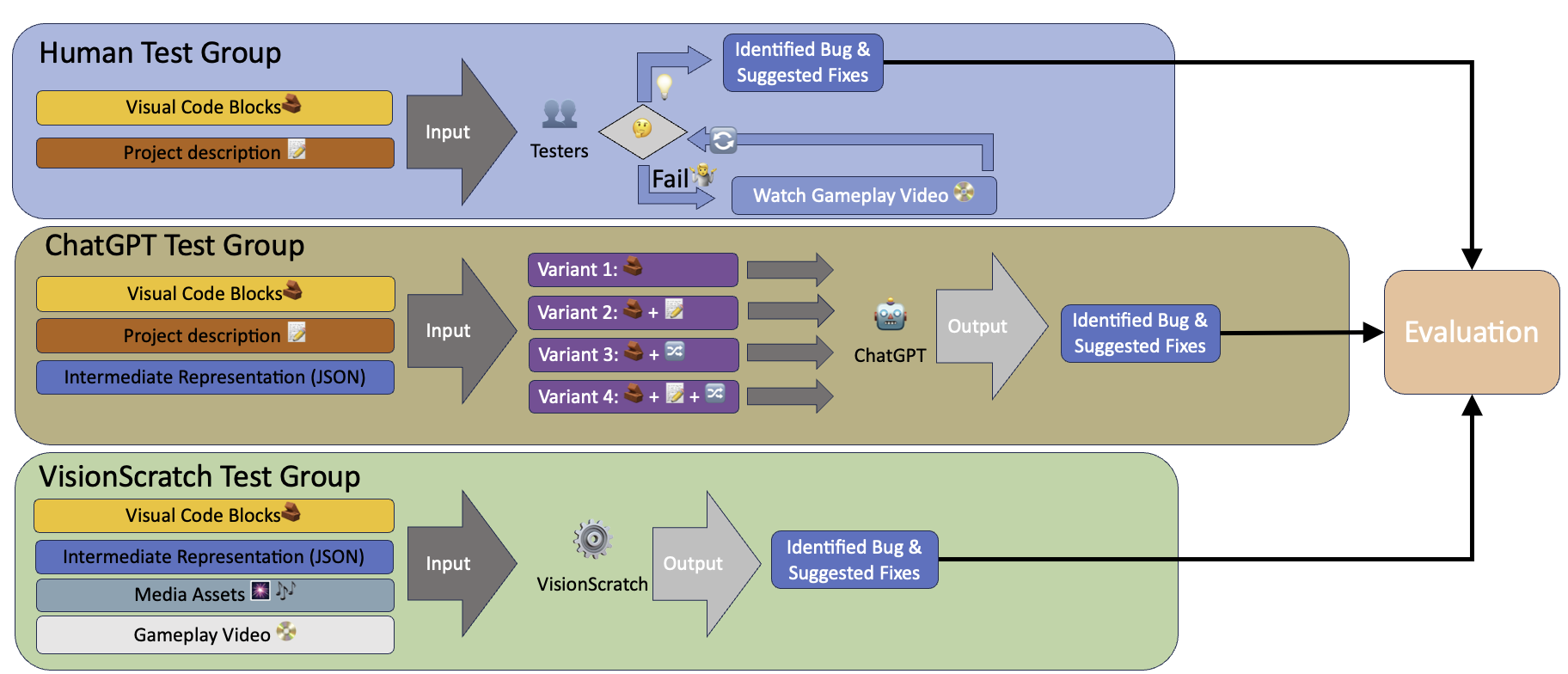}
    \caption{Experiment Procedure for each project. There are three test groups: human, ChatGPT, and \app. The human participants are provided with project description and the block code first, and if they fail to identify the bug, they will be shown the gameplay video in addition and make a second attempt. For ChatGPT's test group, there are four variants running in isolated sessions, with different combinations of input signals in the prompt: visual code blocks only, visual code blocks + project description, visual code blocks + intermediate JSON representation, and visual code blocks + intermediate JSON representation + project description. The \app test group takes visual code blocks, intermediate JSON representation, media assets, and gameplay video as input, but without the project description. For each project and each test variant, the output is in the format of identified bug description and suggested code fix options. These outputs are recorded and then analyzed using the same evaluation criteria.}
    \label{fig:Experiment_procedure}
\end{figure}

\para{Datasets.}
We curated a dataset of 10 faulty projects from the Scratch official program gallery~\footnote{\url{https://scratch.mit.edu/explore/projects/all}}. Projects were selected based on three criteria: clarity, simplicity, and the presence of exactly one bug which falls into the patterns outlined in Section~\ref{sec:merge}. The tasks span three difficulty levels (easy, medium, hard), determined by factors such as the number of blocks and scripts, the number of sprites, and the complexity of control logic. In our experiments, we observed that the time human testers required to resolve a task correlates positively with its difficulty. Table~\ref{fig:Dataset Descr.} lists the project names, bug descriptions, and difficulty levels.
We have three test groups in our experiments as illustrated in Figure \ref{fig:Experiment_procedure}: Human, ChatGPT, and \app.

\para{Implementation.}
We have built a \app prototype using a mix of Python and open-source software libraries. The code of \app’s implementation consists of approximately 800 lines of Python code, which is 5 to 10 times fewer than typical Scratch repair frameworks in educational settings~\cite{deiner2024nuzzlebug,schweikl2025repurr}.
We set the temperature to 1.0 based on preliminary experiments. We ran all the experiments on a MacBook Air (Apple M3 chip, 8GB RAM).

\para{Human Test Group.}
We recruited ten undergraduate and graduate students, each completing a random subset of roughly five project tasks in one-hour sessions. Participants had programming experience ranging from novice to upper-intermediate. Unlike prior work that evaluated tools with children for pedagogical insights~\cite{deiner2024nuzzlebug,STRIJBOL2024101617}, we focused on adults with prior exposure to programming. This choice establishes a stronger debugging baseline and isolates the role of video signals without confounds from reading ability, attention span, or limited programming background. As a result, our evaluation provides a conservative estimate of \app’s effectiveness relative to informed human reasoning. While a classroom study with novice learners remains an important direction, our focus here is on core debugging feasibility.

\para{ChatGPT Test Group.}
We evaluated OpenAI’s ChatGPT as a state-of-the-art debugging assistant, invoking the GPT-4o model through the official interface~\cite{OpenAI_GPT4o_Docs_2025, OpenAI_GPT4o_SystemCard_2024}. We selected GPT-4o because it was the strongest LLM accessible via stable APIs at the time of evaluation. While alternatives such as Claude 3.5 and open-source LLaMA variants were considered, they did not perform well in the stable JSON-editing capabilities needed for Scratch repair and performed less reliably in our evaluation. Our baseline coverage therefore reflects the state of practice in multimodal debugging, rather than an exhaustive survey of all available LLMs.

We designed four prompt variants, each providing different levels of contextual information about the buggy program. Media assets (e.g., \texttt{.svg}, \texttt{.wav}) and gameplay videos were excluded, since GPT-4o does not support these input modalities\footnote{\url{https://help.openai.com/en/articles/7031512-audio-api-faq}}\footnote{\url{https://help.openai.com/en/articles/8400551-chatgpt-image-inputs-faq}}.
\begin{itemize}
\item \textbf{Visual Code Blocks only:} The prompt contained only the program’s block code (as PNG images).
\item \textbf{Visual Code Blocks + Project Description:} In addition to block code, the prompt included a short description of the program’s intended behavior.
\item \textbf{Visual Code Blocks + Intermediate Representation (JSON):} The prompt included block code plus a structural specification of stage, sprite, event, variable, and list interactions.
\item \textbf{Visual Code Blocks + Intermediate Representation (JSON) + Project Description:} This variant combined all available inputs, offering ChatGPT the most comprehensive debugging context.
\end{itemize}
For each variant, the model was instructed to produce a structured output consisting of the \emph{Identified Bug} and corresponding \emph{Suggested Fixes}.

\para{\app Test Group.}
The \app test group evaluated our automated debugging system. Provided with both the gameplay video and the project archive (including visual code blocks, intermediate representation in JSON, and media assets), the reasoning module generated structured outputs consisting of the identified bug and suggested fixes.

\para{Experiment Procedure.}
Each of the three groups described above followed a specific procedure for the debugging tasks, as described in Figure~\ref{fig:Experiment_procedure}. 
\begin{itemize}
    \item \textbf{Human test group:} Participants first attempted to identify the bug using only visual code blocks. If unsuccessful, they were shown the gameplay video and made a second attempt. For each project, we recorded the identified bug and the suggested fixes.
    \item \textbf{ChatGPT test group:} For each project and prompt variant, we ran nine single-turn trials (three trials across three rounds) in isolated sessions. Outputs (Identified Bug + Suggested Fixes) were recorded per trial. A round was considered successful if at least one of the three trials succeeded; average round-level success rates were reported.
    \item \textbf{\app test group:} Followed the same procedure as ChatGPT, using gameplay video and the project archive (blocks, JSON, media assets) as inputs, but excluding the project description.
\end{itemize}

With the recorded outputs from each test group, we manually assessed the results using the same criteria whether the bug was correctly identified and whether the fix suggestions were valid, and then performed data analysis for comparison and evaluation.

Although our dataset includes 10 projects, each was subjected to extensive debugging trials involving ten human participants, four ChatGPT prompt variants, and \app. This design generated more than 200 independent debugging attempts, all manually verified for correctness, yielding a robust basis for comparative evaluation. We in particular followed established precedents in block-based debugging research (e.g., RePurr~\cite{schweikl2025repurr}, NuzzleBug~\cite{deiner2024nuzzlebug}), which relied on similarly curated datasets (e.g., RePurr with 12 projects, NuzzleBug with 8 tasks), stating that high-quality, controlled, user-based evaluations are both credible and impactful at this scale.

\subsection{Results}

\label{sec:results}

We summarize the experimental results in Table~\ref{tab:exp-results}, reporting success rates computed as described in Section~\ref{sec:exp-setup}. Table~\ref{tab:success-labs} details, for each project, the fix success rates across test groups and input variants. Table~\ref{tab:success-testers} aggregates success rates by participant under two conditions: block code only and block code plus gameplay video. Similarly, Table~\ref{tab:success-ai} reports aggregated success rates for the AI groups and variants. Statistical comparisons across groups and variants were performed using two-sided Fisher’s exact tests with $\alpha = 0.05$.

\begin{table}
\centering
\captionsetup[subtable]{justification=centering}
\caption{Success rates across projects and test groups}
\label{tab:exp-results}
\begin{subtable}[H]{\textwidth}
\centering
\subcaption{Per-project success rates by test group and input variant.}
\label{tab:success-labs}
\begin{tabular}{l|l|l|l|l|l|l|l} \hline
 & \multicolumn{2}{c|}{\textbf{Human}} & \multicolumn{4}{c|}{\textbf{ChatGPT}} & \textbf{\app} \\ \hline
Blocks & {\cellcolor{gray!30}\checkmark} & {\cellcolor{gray!30}\checkmark} & {\cellcolor{gray!30}\checkmark} & {\cellcolor{gray!30}\checkmark} & {\cellcolor{gray!30}\checkmark} & {\cellcolor{gray!30}\checkmark} & {\cellcolor{gray!30}\checkmark}\\ \hline
JSON &  &  &  & {\cellcolor{gray!30}\checkmark} &  & {\cellcolor{gray!30}\checkmark} & {\cellcolor{gray!30}\checkmark}\\ \hline
Description & {\cellcolor{gray!30}\checkmark} & {\cellcolor{gray!30}\checkmark} &  & & {\cellcolor{gray!30}\checkmark}   & {\cellcolor{gray!30}\checkmark} & \\ \hline
Gameplay Video &  & {\cellcolor{gray!30}\checkmark} & &&& & {\cellcolor{gray!30}\checkmark}\\ \hline
Media Assets & &&&&&& {\cellcolor{gray!30}\checkmark}\\ \hline \hline
Project 1 (easy)  & 0\%   & 40\%  & 33\% & 0\%   & 100\% & 0\%   & \textbf{100\%} \\
Project 2 (easy)  & 80\%  & 80\%  & 0\%  & 33\%  & 0\%   & 0\%   & \textbf{100\%} \\
Project 3 (easy)  & 20\%  & 40\%  & 100\%& 100\% & 100\% & 100\% & \textbf{100\%} \\
Project 4 (easy) & 60\%  & 60\%  & 0\%  & 33\%  & 0\%   & 0\%   & \textbf{100\%} \\
Project 5 (easy) & 20\%  & 100\% & 33\% & 0\%   & 0\%   & 0\%   & \textbf{100\%} \\
Project 6 (easy)  & 100\% & 100\% & 0\%  & 0\%   & 100\% & 100\% & \textbf{100\%} \\
Project 7 (medium) & 50\%  & 50\%  & 0\%  & 0\%   & 0\%   & 0\%   & \textbf{100\%} \\
Project 8 (medium) & 20\%  & 80\%  & 33\% & 0\%   & 100\% & 0\%   & \textbf{100\%} \\
Project 9 (hard) & 40\%  & 80\%  & 66\% & 100\% & 66\%  & 66\%  & \textbf{100\%} \\
Project 10 (hard) & 20\%  & 40\%  & 0\%  & 0\%   & 0\%   & 33\%  & \textbf{100\%} \\
\hline
\end{tabular}
\end{subtable}

\vspace{0.9em}

\begin{subtable}[H]{0.49\textwidth}
\centering
\subcaption{Human participants success rates}
\label{tab:success-testers}
\begin{tabular}{l|c|c}
\hline
Tester & Blocks & Blocks + Video \\
\hline
Tester 1  & 60\% & 80\% \\
Tester 2  & 50\% & 75\% \\
Tester 3  & 80\% & 100\% \\
Tester 4  & 60\% & 60\% \\
Tester 5  & 60\% & 80\% \\
Tester 6  & 20\% & 80\% \\
Tester 7  & 0\%  & 40\% \\
Tester 8  & 0\%  & 33\% \\
Tester 9  & 83\% & 100\% \\
Tester 10 & 0\%  & 20\% \\
\hline
\end{tabular}
\end{subtable}
\begin{subtable}[H]{0.5\textwidth}
\centering
\subcaption{AI variants success rates}
\label{tab:success-ai}
\begin{tabular}{l|c}
\hline
Variant   & Succ. rate \\
\hline
ChatGPT (Blocks)                 & 27\% \\
ChatGPT (Blocks + JSON)             & 27\% \\
ChatGPT (Blocks + Descr.)            & 47\% \\
ChatGPT (Blocks + Descr. + JSON)   & 33\% \\
\textbf{\app}                & \textbf{100\%} \\
\hline
\end{tabular}
\end{subtable}
\label{tab:success-crossed}
\end{table}

\para{RQ1: Are the program-generated videos on the platform useful for debugging?}

\para{Analysis:} Table~\ref{tab:success-testers} illustrates the impact of video signals on human debugging performance. Across 10 projects (5 participants each, except 6 for Project 7), success rates improved significantly after participants were shown the gameplay video (risk ratio = 1.66, $p = 0.009$). The average success rate with code blocks alone was 41.2\%, increasing to 68.6\% when gameplay video was included.

At the project level, Table~\ref{tab:success-labs} reveals heterogeneity in the gains from video. Projects 5, 8, and 9 exhibited notable improvements, while Projects 2, 4, and 7 remained relatively flat. These differences suggest that the usefulness of video varies with bug type and program complexity. In contrast, \app consistently achieved strong performance across all tasks, indicating that LLMs may extract and utilize video signals more effectively than humans.

In the ChatGPT test group (Table~\ref{tab:success-labs}), combining visual blocks with JSON representations and project descriptions sometimes led to degraded performance, highlighting the unreliability and noise introduced by these input formats. By contrast, \app’s multimodal design, which integrates gameplay video with structured code artifacts, demonstrates clear advantages in both reliability and effectiveness.

\para{Human participants feedback:} Qualitative feedback aligns with the quantitative findings. Participants reported that gameplay videos made program behaviors more legible and revealed bugs that were otherwise invisible in the code. Many emphasized that seeing sprite behavior helped them quickly locate logic flaws. Overall, testers agreed that the generated videos served as intuitive debugging aids and made the process substantially easier.

In summary, we demonstrate the usefulness of video signals for debugging, both for human participants and \app.

\para{RQ2: How well can \app identify and fix program bugs?}

\para{Analysis:} As shown in Tables~\ref{tab:success-labs} and~\ref{tab:success-ai}, across 10 projects (30 rounds of trials), \app achieved a 100\% success rate, outperforming both the strongest human configuration (Blocks + Project Description + Video) (risk ratio = 1.46, $p = 2.94 \times 10^{-4}$) and the best-performing ChatGPT variant (Blocks + Project Description) (risk ratio = 2.14, $p = 1.94 \times 10^{-6}$). Notably, \app does not have access to the project description outlining intended behaviors, yet it surpasses baselines that rely on this information. This result underscores \app's potential for fully automated feedback generation without manual semantic annotations.

ChatGPT, when provided with code blocks and the project description, occasionally achieves perfect success on certain projects (e.g., 100\% on Projects 1, 3, 6, and 8), but fails entirely on others (e.g., 0\% on Projects 2, 7, and 10), indicating highly unstable performance (Table~\ref{tab:success-labs}). Interestingly, augmenting the input with JSON in addition to code blocks and the description reduced the overall success rate from 47\% to 33\% (risk ratio = 0.71, $p = 0.43$). This suggests that the JSON signals may not be effectively leveraged by ChatGPT—possibly due to their noisy structure or because they divert the model’s attention from more salient information.

In contrast, \app integrates multimodal inputs,  specifically, the buggy code artifacts and gameplay video, through a structured pipeline that leverages the JSON extracted from \texttt{.sb3} files to route greater attention to the video signal. This design enables the reasoning module to localize issues more explicitly while mitigating the effects of possible noisy or redundant signals in the JSON. As a result, \app delivers consistent performance across all projects, with accuracy levels substantially exceeding those of ChatGPT.

\para{Human participants feedback:} Participants consistently praised the clarity and usefulness of \app’s debugging suggestions. Several noted that the tool correctly identified the underlying bug and provided precise, actionable fixes. For example, Tester~4 remarked: “Overall, it’s very good. The bug identification and fixes are both excellent.”
Others emphasized \app’s ability to uncover subtle issues that are difficult to detect manually. Tester~7 added: “Very helpful. I noticed some details that human players wouldn’t normally catch, while the tool clearly identified the bug and fix options.”
These comments suggest that \app generates targeted and practical diagnostics by combining visual cues with structural code information.

\para{Pedagogical value of \app.}  
In addition to its strong performance, \app is both efficient and cost-effective. Running the full pipeline end-to-end takes only 52.2 seconds on average, at a total cost of \$0.1185 USD per Scratch project. For comparison, each human participant in our study was allotted up to 10 minutes per task. This efficiency supports the claim that \app can operate quietly in the background as an automated audit tool. These results demonstrate that \app not only delivers high debugging accuracy, but also maintains practical runtime and cost profiles—enabling real-world scalability for classroom deployment and large-scale learner support without incurring prohibitive resource demands.

In summary, \app achieves outstanding performance in identifying and fixing Scratch program bugs, compared to various human and AI baselines.

\subsection{Ablation study: LLM choice for \app}

To evaluate the impact of the LLM backbone on \app’s performance, we conduct an ablation study by varying the reasoning and repair model while keeping all other components fixed (e.g., prompt design, AST-based editing, VM-based verification). 
Specifically, we replace the default Gemini 2.5 Pro model with two other state-of-the-art multimodal LLMs: Qwen3-Max~\cite{Qwen3_techreport_2025} and GLM-4V-Plus-0111~\cite{GLM4V_Thinking_2025, GLM4VPlus0111_Docs_2025}, and run the complete debugging pipeline on the same set of 10 faulty Scratch projects.

For each project and each LLM variant, we allow up to five independent single-turn repair attempts. Following the \app workflow, each subsequent attempt is triggered only if the previous one fails VM verification. A run is marked as successful if any attempt yields a correct bug identification and a verified fix; otherwise, it is counted as a failure.

For each successfully fixed project, we record the value of $k$ at which the first valid fix is achieved (i.e., pass@1 if success on the first attempt, pass@2 for the second, and so on). The mean pass@k is then computed over successful projects only, lower the better. 

This setup allows us to assess not only whether different LLMs are capable of solving the task, but also how efficiently they converge under identical conditions.

\begin{table*}[t] 
\centering 
\begin{tabular}{@{}rlll@{}} \toprule \textbf{Project} & \textbf{Gemini 2.5 Pro} & \textbf{Qwen3-Max} & \textbf{GLM-4V-Plus-0111} \\ \midrule 1 (easy) & Succ. (pass @3) & Fail (no pass after 5 tries) & Succ. (pass @2) \\ 2 (easy)& Succ. (pass @1) & Succ. (pass @4) & Succ. (pass @1) \\ 3 (easy)& Succ. (pass @1) & Fail (no pass after 5 tries) & Succ. (pass @1) \\ 4 (easy)& Succ. (pass @1) & Fail (no pass after 5 tries) & Succ. (pass @1) \\ 5 (easy)& Succ. (pass @1) & Succ. (pass @4) & Fail (no pass after 5 tries) \\ 6 (easy)& Succ. (pass @1) & Succ. (pass @2) & Fail (no pass after 5 tries) \\ 7 (medium)& Succ. (pass @1) & Succ. (pass @4) & Succ. (pass @1) \\ 8 (medium)& Succ. (pass @1) & Succ. (pass @3) & Succ. (pass @3) \\ 9 (hard)& Succ. (pass @1) & Fail (no pass after 5 tries) & Succ. (pass @4) \\ 10 (hard)& Succ. (pass @2) & Fail (no pass after 5 tries) & Succ. (pass @1) \\ \bottomrule 
\end{tabular}
\caption{Ablation results on the backend LLM used in \app. We evaluate three model variants—Gemini 2.5 Pro, Qwen3-Max, and GLM-4V-Plus-0111—in the reasoning and repair modules, while keeping all other system components unchanged (e.g., prompt design, AST editing, verification). Each run uses the full multimodal input: buggy code blocks, JSON, media assets, and gameplay video. A run is marked as Succ. (pass@k) if the model outputs a correct bug identification and suggested fix, verified within k $\leq$ 5 attempts. If all the five attempts fail verification, the run is marked as Fail (no pass after 5 tries).}

\label{tab:llm-ablation} 
\end{table*}

\para{Results.}  
As shown in Table~\ref{tab:llm-ablation}, Gemini~2.5~Pro successfully repaired all 10/10 projects within three attempts. Specifically, eight were solved on the first attempt (pass@1), one on the second (pass@2), and one on the third (pass@3), yielding a mean pass@k of 1.3 across successful runs.  
GLM-4V-Plus-0111 solved 7/10 projects, with 5 resolved on the first attempt, one on the second, and one on the third (mean pass@k = 1.43).  
Qwen3-Max completed only 2/10 projects, with one success at pass@2 and one at pass@3 (mean pass@k = 2.5).

Key observations include:
\begin{itemize}
    \item Accuracy on first attempt: Gemini 2.5 Pro achieved the highest pass@1 rate (8/10), significantly outperforming GLM-4V-Plus-0111 (5/10) and Qwen3-Max (0/10), indicating superior alignment and patch generation from the outset.
    \item Robustness: Gemini 2.5 Pro exhibited no failures under the attempt budget. In contrast, both alternative models failed on multiple projects and succeeded only at later attempts, indicating fragility.
    \item  Effectiveness: Gemini 2.5 Pro required the fewest attempts per successful fix (1.3), followed by GLM-4V-Plus-0111 (1.43) and Qwen3-Max (2.5), further demonstrating its effectiveness under constrained interaction budgets.
\end{itemize}
Overall, Gemini 2.5 Pro is the most reliable and efficient backbone LLM for \app, delivering state-of-the-art multimodal repair with consistent first-pass success and minimal retries.

%% file: secs/discussion.tex
\section{Threats to Validity}

\para{Internal validity.}
Our findings may be affected by design choices in both evaluation and implementation. Like its underlying LLM, \app is nondeterministic and offers no guarantees of soundness or completeness; we mitigate this by sampling multiple candidate repairs, ranking them, and requesting clarifying learner feedback. Correctness validation relies on video playback, which is weaker than formal test oracles but remains the most authentic and widely accepted debugging surface in Scratch, where correctness is inherently perceptual. Data leakage from pretraining cannot be entirely excluded, although we minimized this risk by using a self-collected dataset from the official Scratch forum rather than public repositories. Our evaluation does not include every recent LLM (e.g., Claude 3.5, LLaMA derivatives), but we mitigated this by selecting representative state-of-the-art systems (GPT-4o, Gemini 2.5 Pro) and running ablations with Qwen3 and GLM-4V. This triangulation yields a diverse and credible baseline set while keeping the study feasible.

\para{External validity.}
The generalizability of our results is constrained by the dataset scope. We evaluated ten curated Scratch projects, comparable to prior block-based repair studies~\cite{schweikl2025repurr}, but not sufficient to cover the full diversity of learner programs. Our system also targets Scratch only. We selected Scratch for its popularity and accessible runtime, while the core design of combining code and video is language agnostic. Scaling is difficult due to the need for gameplay video, manual validation, and user debugging trials. We therefore emphasized depth over breadth, with each project tested by multiple human participants and AI baselines, yielding more than 200 verified attempts. Extending to larger datasets and additional environments is an important avenue for future work.

\para{Construct validity.}
Correctness validation in \app relies on perceptual video playback rather than formal test oracles. While this may appear weaker from a traditional software engineering perspective, in the Scratch domain correctness is inherently visual and judged by what learners see on screen. Classroom practice confirms that teachers and learners routinely validate programs by watching behavior rather than checking against hidden test suites. Thus, our oracle directly reflects authentic educational evaluation criteria: a repair is meaningful if it restores expected visual behavior in the project. Beyond correctness, we further assess pedagogical value by ensuring that generated fixes are minimal and interpretable, so that learners can understand the change rather than receiving a full replacement solution.

%% file: secs/related_work.tex
\section{Related Work}
\para{AI Copilots for Scratch.} Inspired by the success of AI coding assistants in text-based programming (e.g., GitHub Copilot), recent work has developed copilots tailored to block-based environments. ChatScratch ~\cite{chen2024chatscratch} and Cognimates Copilot\footnote{\url{http://cognimatescopilot.com/}} embed LLMs into structured storyboarding tools, helping children develop game or story ideas and generate sprites or backdrops from descriptions or drawings. MindScratch ~\cite{chen2025mindscratch} adapts this paradigm for classroom use, guiding students with an AI-supported mind-mapping interface that links project design to teacher-defined learning goals. These copilots show promise in supporting creativity and planning, but none address the core challenge of debugging: detecting and repairing visually implicit errors in block interactions. \app is the first system to fill this gap, combining video analysis with program repair to deliver debugging support that is both pedagogically grounded and aligned with Scratch’s inherently visual nature. Unlike these copilots that focus on creativity support, \app addresses the overlooked but critical dimension of debugging, where correctness is perceptual and grounded in video.

\para{Debugging Support for Scratch.}
Several tools support learners in diagnosing and repairing Scratch programs. Interactive debuggers such as NuzzleBug~\cite{deiner2024nuzzlebug} and Blink~\cite{STRIJBOL2024101617} add stepping, pausing, breakpoints, and even reverse execution, making program behavior more transparent. RePurr~\cite{schweikl2025repurr} introduces automated repair, evolving candidate patches through genetic programming guided by fault localization. Static analyzers highlight novice issues such as dead code, missing handlers, or unused broadcasts~\cite{boe2013hairball,techapalokul2017qualityhound,moreno2015drscratch,fradrich2020commonbugs,fraser2021litterbox}, while testing frameworks generate inputs or assertions to uncover behavioral errors~\cite{johnson2016itch,stahlbauer2019whisker,deiner2023autotest,goetz2022modelbased}. \app is distinguished by triangulated baselines (human, ChatGPT, and ablations) and the use of video as a debugging signal, enabling stronger comparative analysis at similar scale. Existing tools remain limited: debuggers require extensive user interpretation, search-based repair depends on heuristics, and analyzers or tests are restricted by rules or available specifications. By contrast, \app unifies code and gameplay video as an external specification, detecting subtle semantics errors beyond the reach of symbolic or test-based approaches.

\para{Automated Feedback in Educational Programming.} Automated feedback systems~\cite{ko2004whyline, barr2014plastic,keuning2019automatedfeedback,liffiton2023codehelp,SantolucitoZZCP22} for novices often rely on rule-based hints that flag known error patterns~\cite{ZhangPZX21,AssuageHu} or generate fixes from expert-authored solutions~\cite{0002LKSP22}. In block-based settings, iSnap~\cite{price2017isnap} compared student progress against expert states or policies learned from prior learners~\cite{rivers2017itap}. CATNIP further demonstrated automated hints for Scratch by leveraging instructor-authored test suites to identify behavioral mismatches and suggest edits~\cite{fein2022catnip}. While these approaches improve debugging outcomes, they depend on predefined solutions, curated tests, or large datasets, constraining their use to structured assignments and predictable errors. \app departs from this paradigm by deriving intended behavior directly from gameplay videos, supporting open-ended projects without pre-authored scripts, and offering multiple candidate fixes to preserve learner agency.

\para{Large Language Models for Generating Programming Hints.} Recent work has explored using LLMs such as Codex and GPT-4 for automated debugging, including bug detection, code repair~\cite{legoues2015manybugs,Xia2024,huang2025seeing,PyDex,ZhangMKPL22}, and explanatory feedback~\cite{chen2021codex,phung2023feedback}. While effective in general software engineering, LLM outputs are often unreliable, introducing new errors, producing full rewrites, or offering excessive fixes~\cite{song2023maps,Shao0S00025,Shi00025}. In novice programming education, early studies show promise for LLM-generated hints~\cite{griebl2023blockllm}, but concerns remain around accuracy, pedagogical fit, and hallucinated feedback. \app addresses these challenges by embedding the LLM in a constrained environment: first aligning video with code to identify a single critical bug, then proposing a minimal, verified patch. Our evaluation demonstrates how LLM can be used reliably in open-ended, perceptual domains where naive prompting fails. By checking suggestions against project structure and allowing learners to choose among candidate fixes, \app provides learner-centered debugging support.

%% file: secs/conclusion.tex
\section{Conclusion}

We introduced \app, the first vision–language copilot for Scratch that unifies gameplay video with project code to identify and repair Scratch bugs. Our work provides a new foundation for software engineering in visual programming: correctness in Scratch is fundamentally perceptual, and video offers the most direct means of checking program behavior.
By elevating video to a first-class debugging signal, \app advances beyond rules, test suites, or symbolic traces toward multimodal, perception-aware debugging. More broadly, our findings demonstrate that leveraging video as a specification opens a new avenue for research, enabling multimodal debugging support for block-based languages.